# Normal incidence sound insulation provided by sonic crystal acoustic screen made from rigid scatterers – assessment of different simulation methods.

M.P. Peiró-Torres[1,2]; M. Ferri[2]; L.M. Godinho[3]; P. Amado-Mendes [3]; F.J. Vea Folch[1]; J. Redondo[4]

(1) BECSA. Ciudad del Transporte II. C/ Grecia, 31, Castellón (Spain)

(2) Universitat Politècnica de València. Centro de Tecnologías Físicas, Acústica, Materiales y Astrofísica, División acústica. Camino de Vera s/n, Valencia (Spain).

(3) University of Coimbra. ISISE, Department of Civil Engineering. Rua Luis Reis Santos, 3030-788 Coimbra (Portugal).

(4) Universitat Politècnica de València. Instituto de Investigación para la Gestión Integrada de zonas Costeras, Paranimf 1, Grao de Gandia, Valencia (Spain).

**Abstract**

Sonic crystal acoustic screens have been in progressive research and development in the last two decades as a technical solution for mitigating traffic noise. Their behaviour is quite different from that observed in classical barriers, with the latter being based on physically blocking the direct sound propagation path (only allowing diffracted noise to reach sensible receivers), and sonic crystals providing attenuation efficiency based on the creation of "band-gaps" at specific frequency ranges, due to the Bragg's interference phenomenon. The distinct physical mechanisms of these two types of noise barriers complicates the use of classical simplified or even numerical models developed for traditional barriers to simulate and predict the attenuation performance of a sonic crystal, and alternative methods become thus required. In the acoustics scientific literature, several authors have proposed estimation and simulation methods based on different numerical tools to predict the sound insulation provided by these new noise abatement solutions. This paper presents a comparative evaluation of some of these methods, with emphasis to the assessment of their accuracy versus memory usage in order to determine which one is the most suitable for optimization methodologies in the design of new devices with improved acoustic performance.

**Notations:** BEM: Boundary element method, FEM: Finite element method, FDTD: Finite difference time domain, MFS: Method of fundamental solution, MS: Multiple scattering,

NRD: Noise reducing device, PML: Perfectly matched layer, SCAS: Sonic crystals acoustic screens

Corresponding author: (J. Redondo[4]). fredondo@fis.upv.es

1. Introduction

Noise pollution is a major environmental problem affecting urban areas close to transportation infrastructures; thus, reducing its impact on citizens is an important challenge to be faced. Actions can be taken at both emission or transmission phases; and probably the most used devices to reduce the sound transmission of outdoor noise sources are the acoustic barriers placed between the source and the area to be protected. Since the efficiency of noise reduction by means of barriers depends directly on their height, appropriate implementations are sometimes intrinsically linked to a heavy environmental, urban, visual or aesthetic impact. In recent decades, a solution based on Sonic Crystal Acoustic Screens (SCAS) has been applied to reduce these impacts, with an acceptable acoustic performance. Sonic crystals are defined as heterogeneous media embedded in air, formed by periodic arrangements of acoustic scatterers separated by a predetermined lattice constant [1]. These structures provide a noise control mechanism related with the fact that the multiple sound wave scattering process leads to frequency ranges, called band gaps, in which the wave propagation is restricted [2], as formulated by the Bragg's interference principle. There are several studies that show the application of these concepts in the development of Noise Reducing Devices (NRD) as Sonic Crystal Acoustic Screens (SCAS) [3, 4]. The most recent advances on these devices have been achieved mainly thanks to the application of numerical methods in their design and optimization processes.

The importance of correctly predicting the acoustic performance of new NRDs, even before prototyping them, has led the scientific community to develop and validate several numerical methods that evaluate the acoustic performance of these devices. In fact, these simulation methods have led to the improvement of technology in the field of acoustics and the development of new NRD, as in the case of sonic crystals.

Optimization methodologies applied together with numerical simulation methods are powerful design tools, since they allow devices with improved acoustic performance and new functionalities to be obtained [5]. To optimize the acoustic performance of sonic crystals noise barriers, an accurate and low computational cost simulation method is needed to tackle a complex optimization process that requires multiple iterations and an accurate prediction of the acoustic performance of the different proposed designs. The current work aims to study the relationship between the accuracy of several numerical methods, and their associated computational cost. In order to compare the accuracy of the different methodologies in a simple manner, objective parameters describing the performance of the noise barrier predicted by each method need to be compared, preferably making use of single number descriptors. The standards EN 1793-2 and EN 1793-6 [6, 7] describe test methodologies for measuring the airborne sound insulation of NRDs, depending on whether or not the device will be installed in reverberant areas, and define single valued figures of merit called $DL_R$ or $DL_{SI}$, which weigh the insulation measured in one-third octave bands. In this paper, we will apply a parameter analogous to the above-mentioned "figure of merit", referred as global single-number rating of Sound Insulation index (DL$_{SI}$), whose estimated error will be used to assess the accuracy of the numerical methods under study.

Similarly, the computational cost has been defined in terms of a single numerical value independent of the technical characteristics of the computer equipment used. Thus, the figure of merit applied in this work to evaluate the computational cost is the memory usage. Since reducing simultaneously at each method the two figures of merit -i.e. reducing the estimated error while reducing the memory usage- is not affordable; it can be accepted that we are performing a multi-objective evaluation of each numerical method attending to two antagonistic objectives, accuracy and cost. Therefore, a single answer about which is the perfect method and the optimum values of the control parameters cannot be stated. However, for each particular design, the results discussed in this work shall provide a guide for selecting the most appropriate numerical method and control parameter values.

The paper is developed as follows. First, the numerical methods used in this study will be briefly reviewed, as well as the control parameters of each method directly associated with their accuracy and computational cost. Secondly, the simulation scheme used for all methods will be described and the methodology used to calculate a global airborne sound insulation index for all simulation methods will be presented. Then, the results of the simulations for each of the methods will also be illustrated and a systematic study of the estimated error of global transmission coefficient, $\varepsilon(\tau)$, and the associated memory usage is carried out. Finally, the conclusions and discussions of the study will be presented.

## 2. Simulation Methods Under Study

Several methods have been used in last decades to evaluate the performance of periodic structures in acoustics. One of the first proposed methods was Multiple Scattering (MS). This numerical method simulates the propagation and interaction of wave fields with obstacles. In the classical MS formulation, applied to sound waves interacting with rigid scatterers, the total acoustic field is calculated considering that the field reflected by one obstacle induces further reflected or scattered fields to all the other obstacles, in an iterative manner. In the particular case of cylinders, the reflected field can be obtained analytically, as a result, MS is a semi-analytical method.

In 1913, Zaviska [8] described the MS method for studying the scattering behaviour of finite arrays in 2D acoustic fields. Later, Ignatowsky applied this method to investigate the physical phenomenon of normal incidence in an infinite row of cylinders, in 1914 [9]. Subsequently, several authors [10, 11] presented extensions of those works applied to the case of oblique incidence.

An important parameter affecting the accuracy of this method is the number of iterations or reflections considered in calculations, commonly called the order of the approach. Periodic boundary conditions are not applicable in this method. Therefore, since in this work we are simulating periodic structures, the number of repetitions of the unit cell is taken as a control parameter that affects to both accuracy and memory usage.

As an alternative to semi-analytic methods, there are domain discretization methods, such as Finite Element Method (FEM) or Finite-Difference Time-Domain (FDTD) Method. The FEM method, applied in the present paper is a mesh based method with second-order Lagrangian elements that solves problems by turning a differential problem into an algebraic one by discretizing a continuous medium into several finite elements connected to each other at nodal points. All elements are delimited by sides of other elements or by the contour of the domain. The shape functions define the elemental stiffness matrix of each element which, when assembled, generate the global stiffness matrix. The system of equations is solved by establishing the appropriate boundary conditions, obtaining solutions for each mesh node.

There are many studies that legitimate the use of FEM to evaluate the behaviour of periodic structures. As an example, in [12], FEM has been used to analyse periodic structures and the generation of band-gaps. M. Liu et al used a wavelet-based FEM to investigate the band structure of 1D phononic crystals [13], and, more recently, Sánchez-Perez et al [14] used a 2D FEM model for the design of a SCAS.

In this method, the accuracy and computational cost of calculation is majorly associated with the mesh size. For achieving a reasonable accuracy, this size is conditioned by geometrical parameters, such as the wavelength and the different sizes of the elements.

Another domain-based method is the so-called Finite-Difference Time-Domain (FDTD). This method, originating from electromagnetism [15], was adapted to acoustics about two decades ago [16]. In the case of sound waves in fluids, conservation of momentum and continuity equations are written as two linked update equations for sound pressure and particle velocity, allowing the impulsive response of a system, and therefore its transfer function, to be obtained. The main advantage is that, being a technique that works in the time domain, a single simulation can cover a large frequency range, while its main disadvantage, as in other volumetric methods, is that the memory usage increases enormously when the integration domain is large compared to the wavelength.

The works of Cao et al [17] and Miyashita [18] can be mentioned as precursor works on the use of FDTD for the study of sonic crystals. In the first one, it was demonstrated that this technique allows the band-structure calculations in a very effective way, while the second one is focused in the study of wave guides based on sonic crystals

In the case of FDTD, the accuracy and computational colst will depend almost exclusively on the size of the elements. For the sake of simplicity, other aspects, such as the type of perfectly matched layer (PML), the use of non-cartesian grids or conformal techniques will not be considered here. The Courant number has been set to 1, in order to ensure the stability of the numerical technique.

Differently from the FEM and FDTD, the Boundary Element Method (BEM) is based on the discretization of the boundaries of the analysis domain. Mathematically, the BEM is based on the application of the boundary integral equation at a set of nodes defined along a discretized boundary, allowing for the construction of a system of equations whose solution is the acoustic pressure or the normal particle velocity at these boundary nodes. This method requires a priori knowledge of the Green's function associated with the problem under study.

Some works can also be found regarding the application of the Boundary Element Method (BEM), such as the work of Li et al. [19], in which the BEM is used to perform band-gap calculations of solid sonic crystals, and the work by Koussa et al. [20], in which the BEM is used to study the efficiency of an acoustic barrier complemented by a sonic crystal. Gao et al. [21] also analyzed the band structure using the BEM together with the block SS method. According to the authors, this approach has proven to be effective, allowing the numerical eigenfrequency analyses of periodic phononic structures. An interesting approach was proposed by Karimi et al. [22], who developed a specific BEM algorithm tailored for the analysis of periodic systems, which exploits the periodicity of the geometry to reduce the computational cost.

Luis, please, add a paragraph explaining the main parameters of this method: Number of boundary elements per scatterer, element order, number of nodes by element…(see table 1)

In recent decades, a new class of numerical methods has emerged, namely meshless methods, which have been in progressive development, aiming mostly at reducing computational cost and the effort involved in the discretization of the problem geometry. Within this class, the Method of Fundamental Solutions (MFS) has deserved attention for acoustic problems, since, as happens with the BEM, it makes use of Green's functions that can directly account for infinite or semi-infinite spaces. However, its mathematical formulation and implementation is much simpler, since it is based on a collocation approach without requiring any numerical or analytical integration. In fact, the method is simply based on a linear superposition of fundamental solutions to approximate the solution of the problem, assuming sources located outside of the computational domain to avoid singularities in the solution. There is extensive literature regarding the MFS and its application to acoustic scattering and/or radiation problems, such as the early works of Fairweather et al. [23].

There are only a few examples in the literature regarding the application of the MFS to the study of Sonic Crystals. The first application of the MFS in this field is due to Martins et al. [24], who proposed the use of the MFS to evaluate the sound insulation provided by a periodic structure made of rigid scatterers. Santos et al. [25], extended the formulation to allow considering elastic shell scatterers. However, in both works, the classic formulation of the MFS was used, involving the discretization of all scatterers, and disregarding the periodicity of the structure. More recently, Godinho et al [26] successfully used an improved version of the MFS, developed for finite periodic structures. In Godinho et al [27], the method was further developed to allow accounting for infinite periodic structures along one direction, in a very efficient manner.

3. Simulation Scheme and calculations

As aforementioned, the main aim of this work is to evaluate the suitability of different methods for the simulation of sonic crystal structures. For this assessment, several variations of the same scheme, consisting of sonic crystals structures with a fixed lattice constant, have been simulated by all considered methods, with different radii of the scatterers. The geometry of the studied configuration (Fig 1) consists of a square array of cylindrical scatterers placed in four rows, separated by the lattice constant, a=0.17m, so that the first band gap, usually called Bragg's gap, appears around the frequency of 1000Hz, the most relevant frequency of the normalized traffic noise spectrum, standardized by EN 1793-3 [28]. In order to study the ideal behavior of this sonic crystal neglecting diffraction effects at its edges we should model an infinite three-dimensional volume with infinite scatterers of infinite height. This infinite domain is unaffordable computationally, but in the case of perpendicular plane wave incidence, its mathematical solution is equivalent to that of a simplified two-dimensional (2D) numerical model, consisting of a single cell of the structure where periodic boundary conditions are imposed on both lateral contours of the computational domain (Fig 1). At this single cell, where 2D simulations are performed, we will consider an incident plane wave impinging perpendicularly the screen, and several measurement points located in a square array along the measurement area, in twelve lines parallel to the plane wave-front, separated a/4 from each other, the first of these lines being placed 3a/2 apart from the center of the nearest scatterer. In order to avoid duplication of data, since the unit cell is symmetrical with respect to an axis perpendicular to the plane wave-front passing through the center of the scatterers, the measurement points are placed between this symmetry axis and one lateral boundary. It must be said that periodic boundary conditions can be applied for all methods except MS, in fact, for this method, the number of repetitions of the 2D unit cell is considered a control parameter affecting the accuracy of the results (Table 1)

Several scatterer diameters were tested. For the sake of brevity, we present here only three representative cases. The results will be shown for each of these three diameter values, expressed as a fraction of the lattice constant, 0.25a, 0.5a, 0.75a.

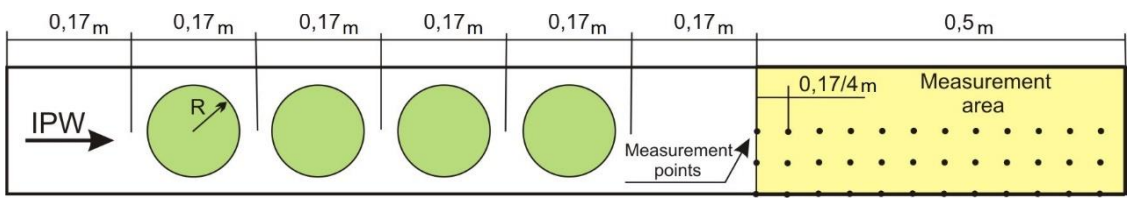

Fig. 1. Schematic representation of the system configuration used in the simulations.

With the aim of obtaining a single figure of merit to quantify the acoustic performance of the evaluated devices, the Sound Insulation index ($DL_{SI}$) was calculated, based on the standard EN 1793-6 [7].

$$DL_{SI} = -10 \log \frac{\sum_{j=1}^{18} 10^{0,1L_j} 10^{-0,1SI_j}}{\sum_{j=1}^{18} 10^{0,1L_j}} \qquad (7)$$

Being $L_j$ the normalized A weighted sound pressure level, in decibels, of traffic noise in the j[th] one-third octave band defined in EN 1793-3 [28] and $SI_j$ the average Sound

Insulation index at the measurement points in the j[th] one-third octave band of the frequency range of interest (100-5000 Hz)

In other words, $DL_{SI}$ was obtained from a global transmission coefficient, $\tau$, namely:

$$DL_{SI} = -10 \log \tau \quad (8)$$

being $\tau$ obtained as a weighted average of the j-th third octave values of the transmission coefficient, $\tau_j$.

$$\tau = \sum_{j=1}^{18} C_j \tau_j \quad (9)$$

where the coefficients $C_j$ express the normalized traffic noise spectrum with A-weighting, obtained from its normalised one-third octave band levels, $L_j$, by the expression:

$$C_j = \frac{10^{0,1L_j}}{\sum_{j=1}^{18} 10^{0,1L_j}} \quad (10)$$

And analogously, transmission coefficients of the j-th third octave band ($\tau_j$) are related with the Sound Insulation index ($SI_j$) by

$$\tau_j = 10^{-0.1 SI_j} \quad (11)$$

In time domain method (FDTD), to calculate $\tau_j$ at each measurement point, the Fourier transform of the impulse response is obtained and averaged in one third octave bands. In frequency methods, a number of frequencies for each one third octave band, separated a constant octave fraction between them, are evaluated and $\tau_j$ is obtained by averaging the results of all the frequencies inside a band. Finally, $\tau_j$ is averaged for all the measurement points.

To evaluate the accuracy of the calculations, the indicator used has been the estimated error of the global transmission coefficient, $\varepsilon(\tau)$. For this estimation we need to obtain the values of convergence of the transmission coefficients per third octave band, $\tau_{j\_CONV}$, as will be described in the results section. Then the estimated error of the global transmission coefficient is obtained as

$$\varepsilon(\tau) = \sqrt{\sum_{j=1}^{18} C_j \left(\tau_j - \tau_{j\_CONV}\right)^2} \quad (12)$$

Just like this indicator, $\varepsilon(\tau)$, quantifies the concept of "accuracy", the concept "computational cost" will be quantified in this work by the memory usage. The computational cost is a common way of evaluating the efficiency of simulations [30, 31], but it can be associated with two concepts, (i) the memory usage or (ii) the computational time, CPU time, or number of operations, which are highly dependent on the type of implementation and hardware. In this work, therefore, the authors have opted for the memory usage to evaluate the computational cost of each simulation.

4. Preliminary studies.

To properly calculate the estimated error, $\varepsilon(\tau)$, necessary to quantify the accuracy, we need to define a criterion for the calculation of the convergence transmission coefficients, $\tau_{j\_CONV}$, applied at equation (12). With this aim we have performed, for each method, several preliminary series of simulations observing that the dispersion plots between the control parameters vs sound insulation, $DL_{SI}$, showed a convergence value of $DL_{SI}$ slightly different between methods, as shown in Table 2. Given this difference between methods, we have defined the convergence value of $DL_{SI}$ as that with an error two orders of magnitude smaller than the error between methods. I.e., our series of simulations finish when the dispersion of a set of three simulations with consecutive values of the control parameter is smaller than one hundredth of the average error between methods. Then, the average value of the convergence transmission coefficients, $\tau_{j\_CONV}$, obtained from these last three values of the series is taken to evaluate the estimated error of any simulation performed at preliminary studies, leading to define the set of control parameters (Fig 2).

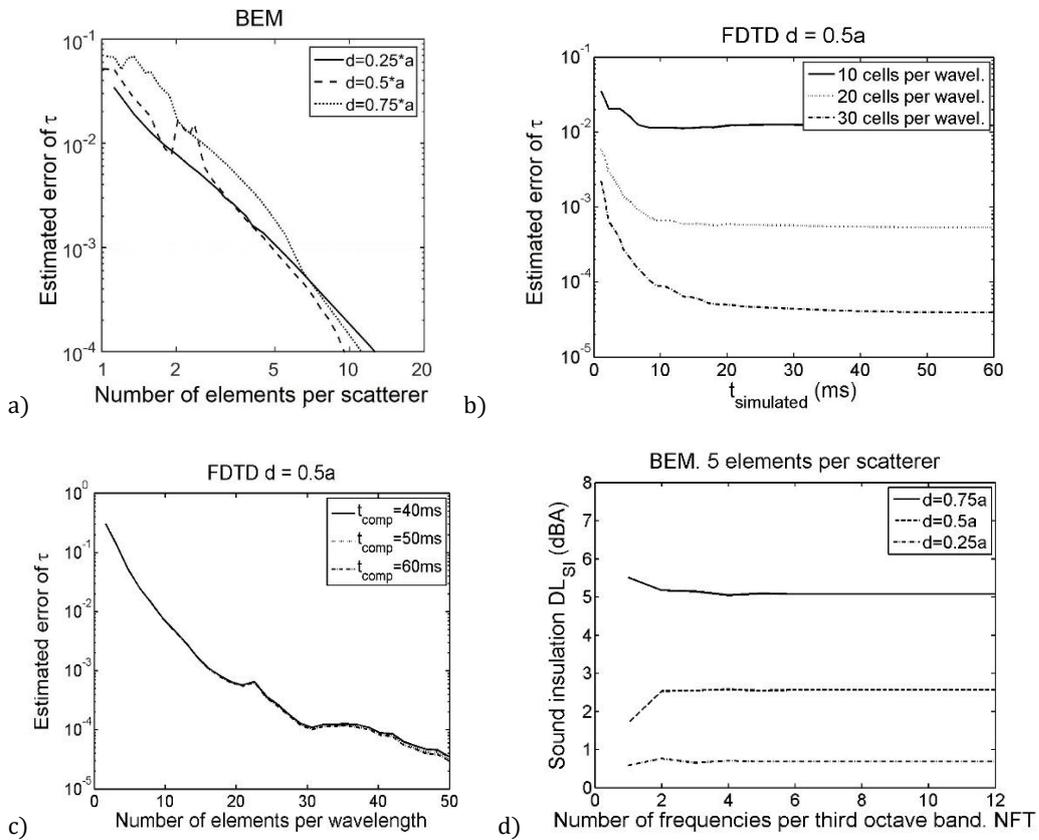

Fig. 2. Evolution of the estimated error of the global transmission coefficient (a) in BEM as a function or the number of elements per scatterer, and in FDTD as a function of (b) the time simulated and (c) the number of elements per wavelength (d) the number of frequencies per third octave band (NFT).

From the evidence of the preliminary studies, we can differentiate between two types of control parameters according to their effect on memory usage and accuracy. Some parameters affect the accuracy only in a range, so we can define for them a saturation value, whereas other parameters affect the accuracy and memory usage in its whole range. We will name these second type of parameters as main control parameters. The additional parameters are kept constant for the rest of the paper. Figure 2a represents the accuracy as a function of a control parameter, such as the number of elements per scatterer in BEM. It can be appreciated that any decrease of this control parameter leads always to a better accuracy; on the contrary, there are other parameters, such as simulation time in FDTD – shown in Fig. 2(b)- whose variations in an interval generate an effect on accuracy, but their effect saturates, with no benefit generated by modifying their value above the saturation value. This example is easy to understand, since if we perform the FDTD in a time so short that the inner reflections between the elements of the sound screen are still taking place, the lectures of the microphones are still uncompleted; but once all the possible inner reflections have been completed, it makes no sense to consider longer times. On the contrary, the mesh size in a volumetric method always will affect the quality of the result, since discretization is just an approach to solve differential equations in a continuous media (Fig 2(a) and Fig 2(c)). Bearing this in mind, we have centered the study in the control parameters with influence, at any range, in both memory usage and accuracy. By the other hand, the constant value chosen for the rest control parameters, such as time in FDTD or reflection order in MS, has been notably oversized with the unique exception of NFT in frequency domain methods, Fig 2(d). This control parameter has not been evaluated at saturation neither in all the range. The reason is that its effect is highly similar for both four frequential methods and, additionally, its value does not affect the memory usage, just the total computation time.

Summarizing, the main control parameters considered in the study, reduced to only one per method are: (i) element size for FEM, FDTD and BEM, (ii) number of repetitions of the unit cell for MS and (iii) number of virtual sources in MFS. Regarding the additional parameters, the number of frequencies for each one-third octave band was set at 6, similarly, the FDTD simulation time was set to 20 times the time required by the sound to cross perpendicularly the sonic crystal. It is noteworthy that, although these two control parameters affect the computational time, they do not significantly affect the computational cost as estimated in the present work. In the case of MS, the number of terms in the Bessel function was set at 5 on the recommendation of several authors [29], and the reflection order set at 3. The main control parameters and the additional parameters are listed in Table 1.

Finally, it must be stated that computational cost, represented here by memory usage, is dependent not only in the control parameters of the method but also in the particular implementations. For this work all the simulations are carried by self-made codes running on Matlab™. All the implementations are basic ones, not including optimization of memory usage, parallelization or GPUs processing.

| Method | Main parameter | Limit values | Additional parameters | Time/freq |
|---|---|---|---|---|
| MS | Repetitions of the unit cell | 1 to 64 | Reflection order=3  Number of terms in the Bessel functions=5 | nft=6 |
| FEM | Mesh quality | Min. number of elements per wavelength*= ¼ to 4  Max. number of elements per wavelength*= 2 to 2000  Growth factor=1.8 to 1.1  Curvature factor 0.8 to 0.2 | Element order. Triangles, i.e. 3 nodes | nft=6 |
| FDTD | Number of cell points per wavelength | 1 to 50 | Courant–Friedrichs–Lewy number = 1 | Total time of flight/time of flight of the 4 cells=20; |
| BEM | Number of boundary elements per scatterer | 1 to 100 | Element order. Constant elements, with 1 central node. | nft=6 |
| MFS | Number of virtual sources and collocation points per scatterer | 1 to 100 | Distance form virtual source to boundary: 40% of the radius of the scatterer | nft=6 |

Table 1. Numerical values of the main parameter considered in each of the simulation methods, as well as the secondary parameters. *Elements or cell points per wavelength refers to the smallest wavelength evaluated ($4000\sqrt{2}$Hz)

## 5. Results and Discussion

The convergence value of the sound insulation index ($DL_{SI}$) is shown at Table 2. A difference of about hundredths of a decibel can be appreciated between methods. The $DL_{SI}$ error in this table has been evaluated by error propagation from the errors of $\tau_j$, estimated as the difference between $\tau_{j\_CONV}$ for each single method and the average, between the five tested method of the $\tau_{j\_CONV}$.

| $DL_{SI}$ (dB(A)) | $d = 0.25a$ | $d = 0.50a$ | $d = 0.75a$ |
|---|---|---|---|
| MS | 0,63±0,03 | 2,45±0,04 | 5,07±0,17 |
| FEM | 0,73±0,06 | 2,60±0,08 | 5,06±0,09 |
| FDTD | 0,66±0,04 | 2,42±0,11 | 5,15±0,22 |
| BEM | 0,679±0,018 | 2,56±0,04 | 5,07±0,08 |
| MFS | 0,678±0,016 | 2,49±0,05 | 5,01±0,05 |

Table 2. $DL_{SI}$ values calculated with the limit values of the control parameters for each numerical method and for each of the diameters referred to.

For a fast inspection of the differences between methods found, shown in table 2, the sound insulation index as a function of frequency ($SI_j$) is plotted in figure 3 for each method. Small differences are observed for high values of the diameter of the scatterers in the first band gap (1kHz) for the MS method. In this case, the sound insulation is slightly underestimated. On the other hand, more evident differences appear in the second band gap (around 2kHz), particularly for the larger scatterers ($d = 0.75\,a$). The underestimation of this second band gap by MS is especially noticeable.

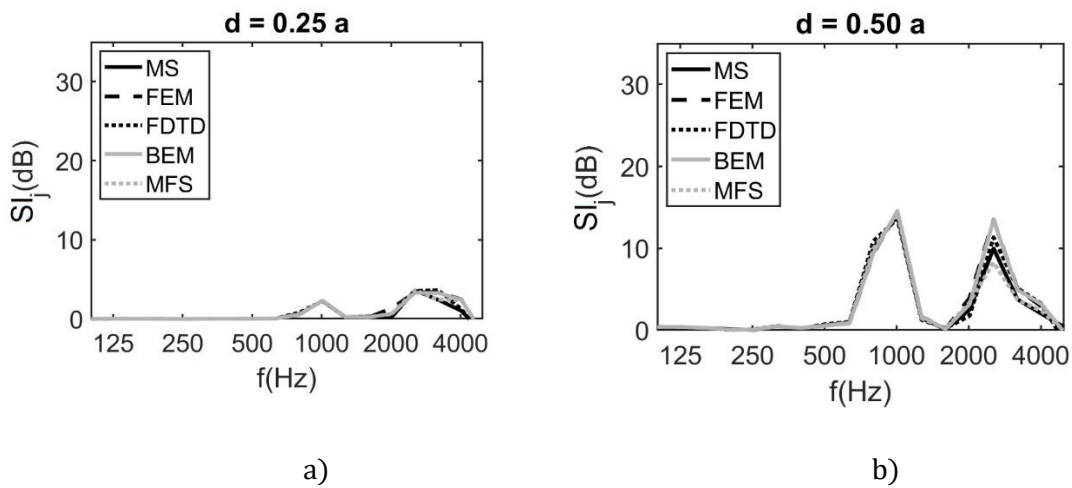

a)          b)

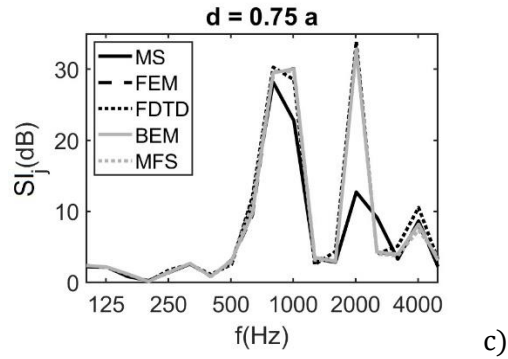

Fig. 3. $SI_j$ vs frequency, in one-third octave bands, for all the considered methods: a) d = 0.25 a; b) d = 0.5 a; c) d = 0.75 a.

As stated in the introduction, the evaluation of the five numerical methods considered is based on the knowledge of the estimated error of the global transmission coefficient as a function of the memory usage involved. And the variations of both, memory usage and estimated error is defined for a particular range of variation of the main control parameter associated to each method. For the estimated error of sound insulation in this final evaluation, we use $\tau_{j\_CONV}$ obtained as the average of the convergence transmission coefficients of the five methods. Figure 4 represents the estimated errors versus the memory usage for each diameter and method considered. As expected, the higher the memory usage, the lower the estimated error. This trend is broken in the case of data with higher memory usage, since the reference value for the error estimation has been obtained as the average of all the methods, so that no error smaller than the estimated error between methods can be found. For the highest memory usage, the estimated error values achieved are around one hundredth of a decibel.

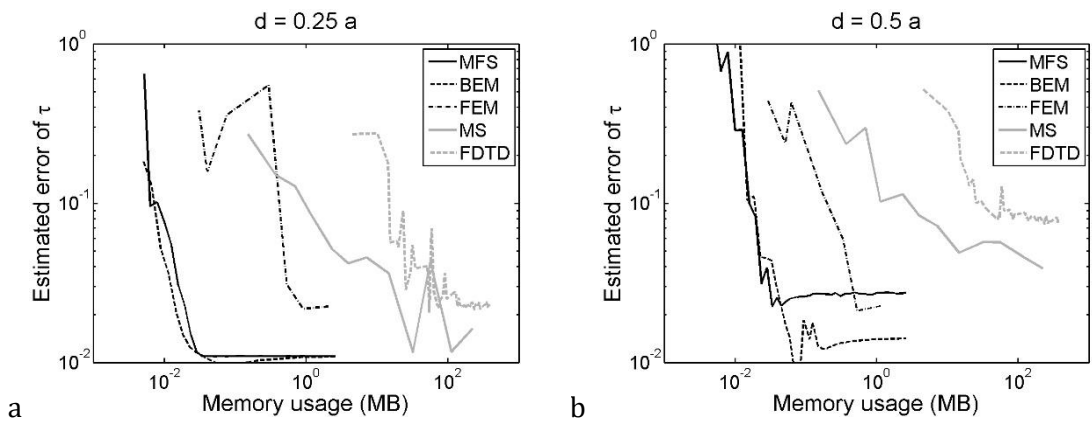

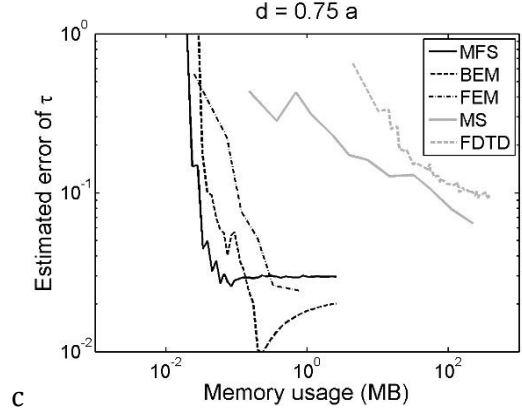

c

Fig. 4. Estimated error of DL$_{SI}$ versus memory usage for all the studied methods: a) d = 0.25 a; b) d = 0.5 a; c) d = 0.75 a.

As an objective evaluation criterion, the best results are those that offer the least estimated error involving lowest memory usage. In other words, representing the estimated error as a function of the memory usage, the best methods are those whose curves are closest to both axes. In some cases, the curve represented by one method may cross with the curve of another method, so there is no absolute preference between them, and the best option would depend on the aims of a particular project.

In a first inspection of the three graphs in Figure 4, there are clear performance differences between the studied methods, according to the evaluation criteria described. BEM and MFS seem to show a much faster convergence than the rest, and FEM presents more favorable results than FDTD and MS. Furthermore, for larger scatterer diameters, FEM shows no significant differences with MFS or BEM. Indeed, for that case, the total number of elements of the BEM (or collocation points in MFS) required to discretize each scatterer is larger, and thus leads to a larger memory usage. By contrast, for these larger diameter scatterers, domain discretization methods such as FEM or FDTD benefit from a small reduction in mesh size (due to the larger void in the mesh corresponding to the scatterers), and thus have improved performance.

A deeper analysis shows that BEM and MFS methods offer very similar behaviour, as can be seen in the three graphs of figure 4 and the convergence value of $DL_{SI}$ (Table 2). Furthermore, in the case of $d$=0.25 there is almost an overlap of BEM and MFS curves, whereas in $d$=0.5 and $d$=0.75 this overlap is only for small memory usages. Although this difference could seem to indicate an irregular behaviour of one of these two methods, it is indeed due to the fact that the reference value for the error estimation has been obtained as an average from several methods. Thus, the method whose convergence value of $DL_{SI}$ is closer to the average of the values of convergence will present a curve that converges better. In fact, for all methods, the shape of the curve depends on the difference between the convergence $DL_{SI}$ value of each method and the average of values of convergence, so their accuracy is affected by the inaccuracy of the other methods.

It is also interesting to note that FEM performs better when the scatterers diameter is larger, as the graph for $d = 0.75$ in Figure 4c shows. This may be due to the way the mesh is defined. As it is the usual practice, a triangular flat mesh has been used, which adapts to the geometry and has a growth function in order to achieve the desired average size, being larger size in the areas far from the scatterers and smaller in areas close to them (see figure 5). Thus, the size gradient will be more abrupt in $d = 0.25$ than in $d = 0.75$ leading to worse behaviour for smaller radius, either because the mesh variability is an additional difficulty for the calculation or simply because of the excess memory usage needed to create the finer meshes surrounding the smaller obstacles. For this reason, the FEM curve of $d=0.25$, not only presents the worst performance with respect to its counterparts of other diameters, but it is also the one that presents a more irregular behaviour, with some increases in estimated error when rising memory usages, in low memory usages ranges. However, as the memory usage increases, i.e., the size of the mesh elements decreases, the representation of the elements becomes more reliable and the results present fewer estimated errors.

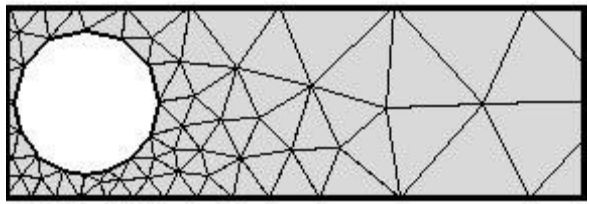

Fig. 5 FEM triangular flat mesh with a high growth factor adapted to the scatterers geometry

Thus, FEM, as a domain discretization method, varies the quality of its calculation according to the way in which the mesh elements are defined to adapt them to the domain to be simulated. However, this is not reflected in other methods. In the case of FDTD, the curve that defines its estimated error as a function of the memory usage, is shifted in the axis of the abscissa about 2-3 orders of magnitude with respect to the methods that give better results (BEM and MFS). This could be due to the fact that the mesh is Cartesian. Therefore, a large mesh size implies a poor definition of the shape of the scatterers, and a poor treatment of the wave dynamics, especially of its high frequencies. But since the method does not make an extra adaptation of the mesh to the geometry of the scatterers, its curve is not affected by the size of the scatterers.

Regarding the difference of several orders of magnitude between the memory usage of the FDTD and that of the MFS, BEM or FEM, it does not imply such a large decrease in the quality of the method, as might be apparent when observing the presented plots. It may be noted that frequency methods use the memory space multiple times, as many times as frequencies to be considered. In FDTD, on the other hand, a unique time domain simulation is made and then, by means of a Fourier transform, the transfer function is obtained. In our case, since we have taken 6 frequencies per each of the 18 one-third octave bands, the number of repetitions performed by the frequency methods is 108.

Finally, MS shows the worse curves between frequency methods, as well as the most discrepant values in the sound insulation curves at convergence, shown in Figure 3. These evidences do not directly rule out the method, it simply evidences that it is the least adapted to periodic structures. In fact, it is the only method in which we could not use periodic conditions.

6. Conclusions

The aim of the present work is to help the researchers in the task of finding a simulation tool that provides maximum precision at the lowest memory usage for the simulation of the acoustic performance of periodic structures. It is essential to have a numerical method with such characteristics when carrying out an optimization process, as this process involves a large number of simulations.

For the particular case raised in the study, both BEM and MFS are the best methods for performing optimization processes and determining the acoustic performance of these periodic structures. Comparing the results offered by FDTD and FEM, both volumetric methods but with different calculation philosophy -since one is based on time domain and the other on frequency domain- we appreciate that FEM effectively gives more accurate results requiring less computational cost, but as explained above, this may be due to the way in which this computational cost is determined, and because we do not take into account to determine the computational cost of the calculation other factors such as computational time.

The option of using calculation time to evaluate the performance of methods was discarded because the calculation time depends strongly on the particular implementation of the method and the particular computer on which it is run. For further research, it is proposed to evaluate not only the memory usage, but also the computational time. Shorter calculation times would shorten the iteration time in optimization processes, leading to greater efficiency of these processes and allowing them to be used as competitive design tools. It could also be interesting to include in the simulation processes absorbent materials and resonant cavities in the acoustic scatterers, which will improve the acoustic performance of the device.

It could be premature to extrapolate the results obtained and discussed in this work to other geometries since each of the methods considered has its own peculiarities, and therefore can be better adapted to particularities such as resonant cavities or absorbents. This may make one method or another more suitable in other situations that have not been considered here. It should be noted that only FEM and FDTD methods are able to directly simulate media with heterogeneities or non-linear media, as they are the only mesh methods that are also volumetric methods.

Further research could be proposed for more complex structures, such as a more complex evaluation of the computational cost, taking into account time but fixing the type of hardware or software applied. This approach would be the subject of a separate research, in this paper we have tried to take a preliminary step in the right direction.

### Acknowledgments

M.P.P.T is grateful for the support of pre-doctoral Grant by the "Ministerio de Ciencia, Innovación y Universidades. Agencia Estatal de Investigación" of Spain through reference nº DI-15-08100

This work has been supported by the Ministerio de Ciencia, Innovación y Universidades, Spain, under grant RTI2018-096904-B-I00.

Support by FEDER funds through the Competitivity Factors Operational Programme - COMPETE and by national funds through FCT – Foundation for Science and Technology of Portugal within the scope of the project POCI-01-0145-FEDER-007633 is also acknowledged.